\documentclass[12pt]{article}
\usepackage{amsmath,amssymb,amsfonts}
\usepackage{longtable}

\def\mylist{\begin{list}{}{\setlength{\leftmargin}{0.5in}
               \setlength{\listparindent}{-0.5in}
               \setlength{\itemindent}{\listparindent}}}

\newcommand{\bra}[1]{\langle#1\,|}
\newcommand{\ket}[1]{|#1\,\rangle}

\newcommand{\nD}[1]{\not D}
\begin{document}

\begin{titlepage}

\title{A comment on the enhancement of global symmetries in superconformal $SU(2)$ gauge theories in 5D}
\author{Denis Bashkirov\\ {\it\small Perimeter Institute for Theoretical Physics, Waterloo, ON N2L 2Y5, Canada\footnote{E-mail: dbashkirov@perimeterinstitute.ca}}}

\maketitle

\abstract{Recently, the superconformal index for a class of 5-dimensional superconformal quantum field theories with gauge group $SU(2)$ and $N_f$ flavors was computed in \cite{KKL} to a few low orders in the chemical potential and enhancement of the global symmetry to $E_{N_f+1}$ was confirmed to this low order. In this note I provide an argument that the information contained in these few low order terms in the expansion of the index is sufficient to prove the enhancement of global symmetries in the theories. So, in particular the expression for the indices must display the enhancement to all orders in the chemical potential.}

\end{titlepage}

\section*{}

 Some time ago Seiberg argued \cite{Seiberg} that five-dimensional ${\mathcal N}=1$ supersymmetric gauge theories with gauge group $SU(2)$ and $N_f$ flavors in the fundamental representation despite being nonrenormalizable by classical power counting posses UV fixed points which are five-dimensional superconformal quantum field theories for $N_f<8$. Formally, the UV superconformal field theories are obtained by setting the dimensionful gauge coupling constant $g$ in the Lagrangian to infinity, that is, by getting rid of gauge kinetic term. The global symmetry seen in the Lagrangian is $SO(2N_f)\times U(1)_I$ where the $U(1)_I$ symmetry is generated by the topological current
 \begin{equation}
 j=\star Tr(F\wedge F)
 \end{equation}
 where $F$ is the gauge field strength two-form and $\star$ is the Hodge-star conjugation. Integral of $\star j$ over a four-cycle computes the second Chern class of the gauge bundle which is called the instanton number due to its relation with four-dimensional instantons. Hence the subscript $I$.

 Using a realization of this gauge theory in string theory Seiberg argued that the global symmetry $SO(2N_f)\times U(1)_I$ is enhanced to exceptional $E_{N_f+1}$. This means that there are additional conserved currents in the theory with appropriate quantum numbers so that taken together with the currents generating $SO(2N_f)\times U(1)_I$ they form the current algebra of $E_{N_f+1}$. The groups $E_{N_f+1}$ for $N_f=0,..,5$ are

 \begin{eqnarray}
 & E_1=SU(2), \quad E_2=SU(2)\times U(1),\quad E_3=SU(3)\times SU(2),\nonumber\\
 & E_4=SU(5),\quad E_5=Spin(10)
 \end{eqnarray}

 Because these additional currents are nonperturbative (all gauge-invariant operators in the theory composed of the fields in the Lagrangian are neutral under $U(1)_I$), the enhancement of the global symmetry can be seen only by some quantities which contain nonperturbative information about the theory. Two of such quantities are superconformal index of the theory put on ${\mathbb S}^4\times{\mathbb R}$ and the partition function of the theory on ${\mathbb S}^5$. Because of the usual operator-states correspondence of CFT's, in this note we are interested in the superconformal index.

 By the operator-state correspondence, there is a one-to-one map between local operators on ${\mathbb R}^5$ and the states of the theory put conformally on ${\mathbb S}^4\times{\mathbb R}$. Some nonperturbative information about the spectrum of states of the theory on ${\mathbb S}^4\times{\mathbb R}$ is captured by the superconformal index \cite{BBMR}. See also \cite{KKL}.
 \begin{equation}
 {\mathcal I}(x, y,m_i,q)=Tr[(-1)^Fe^{-\beta\{Q,Q^\dagger\}}x^{2(R+j_1)}y^{2j_2}e^{-\sum_iH_im_i}q^k]
 \end{equation}

 Here $F$ is the fermion number operator, $R$ is the Cartan generator of the $R$-symmetry $SU(2)_R$, $j_1$ and $j_2$ are third components of the spins of the rotation subgroup $Spin(4)=SU(2)\times SU(2)\subset Spin(5)$, $H_i$ are Cartan generators of the global symmetry $SO(2N_f)$ and, finally, $k$ is the generator of $U(1)_I$. This particular form of the expression for the index is determined by the choice of a supercharge $Q$. The nontrivial contribution to the index comes only from BPS states which are annihilated by $Q$ and by $Q^\dagger$. Because of the anticommutation relation $\{Q,Q^\dagger\}=E-3R-2j_1$ they have quantum numbers $E=3R+2j_1$ where $E$ is the energy in the radial quantization equal to the conformal dimension of the corresponding local operator on ${\mathbb R}^5$.

 Authors of \cite{KKL} applied the localization technique to the path integral of the $SU(2)$ gauge theories with $N_f=1,...,7$ flavors on ${\mathbb S}^4\times{\mathbb S}^1$ to compute the index for several low orders in $x$, $y$, $e^{m_i}$ and the zeroth and first order in $q$. They found the following general pattern
\begin{equation}
 {\mathcal I}_{N_f}=1+\chi_{Adj}^{E_{N_f+1}}x^2+{\mathcal O}(x^3),
\end{equation}
 where $\chi_{Adj}^{E_{N_f+1}}$ is the function of $e^{m_i}$ and $q$ which organizes itself into the formula for the characters of the group $E_{N_f+1}$. This is with the exception of cases $N_f=6$ and $N_f=7$ -- $\chi_{Adj}^{E_{7}}(q)$ and $\chi_{Adj}^{E_{8}}(q)$ contain terms with $q^{\pm2}$, and the corresponding two-instanton contribution was not computed in \cite{KKL} due to technical difficulties.

Now we show that the appearance of the adjoint character for $E_{N_f+1}$ in all cases is not a coincidence and prove the enhancement of the global symmetry. The proof goes in two steps. First, we show that the expression for the index imply the existence of scalars in the triplet representation of $SU(2)_R$, energy (conformal dimension) $E=3$ and the representation of the manifest symmetry group $SO(2N_f)\times U(1)_I$ which corresponds to the decomposition of adjoint of $E_{N_f+1}$ under $SO(2N_f)\times U(1)_I\subset E_{N_f+1}$. Then we show that these scalars are superconformal primary in the supermultiplet of conserved currents.

First of all, both $R$ and $j_1$ in the exponent of $x^{2(R+j_1)}$ are nonnegative. More precisely, the nontrivial contribution comes only from states with nonnegative $R$ and $j_1$. Indeed, suppose $j_1$ is negative. Then take a supercharge $Q'$ with the value of $j_1$ opposite to that of $Q$. In this case $\{Q',(Q')^\dagger\}=E-3R+2j_1$, thus making the matrix element $\bra{BPS}\{Q',(Q')^\dagger\}\ket{BPS}=E-3R+2j_1=4j_1<0$ because for the BPS state $E=3R+2j_1$. By the analogous logic $R$ is nonnegative.

Now, only states with $R+j_1=1$ contribute to the coefficient of $x^2$. The possibilities are: 1) $R=0, j_1=1$ and hence $E=2$, 2) $R=1/2, j_1=1/2$ and hence $E=5/2$, and 3) $R=1, j_1=0$ and hence $E=3$. It turns out that first two options cannot be realized because of the unitarity constraints. Here I list a set of bounds on the quantum numbers of superconformal primary states worked out in \cite{Minwalla}. See also \cite{BBMR}. There are several options depending on the spins of the superconfomal primary states. It is convenient to introduce Cartan generators of $Spin(5)$ as $h_1=j_1+j_2$ and $h_2=j_1-j_2$. An irreducible representation of $Spin(5)$ is labeled by the highest weight $(h_1,h_2)$ with $h_1\ge h_2
\ge0$. Also, in what follows $R$ is the highest weight of $SU(2)_R$, so it is nonnegative.
\begin{itemize}
\item If $h_2>0$ then
\end{itemize}
\begin{eqnarray}
E\ge 3R+h_1+h_2+4
\end{eqnarray}

\begin{itemize}
\item If $h_2=0, h_1\ne 0$ then
\end{itemize}
\begin{eqnarray}
& E\ge 3R+h_1+4\\
&\hbox{or}\quad E=3R+h_1+3
\end{eqnarray}

\begin{itemize}
\item If $h_1=h_2=0$ then
\end{itemize}
\begin{eqnarray}
& E\ge 3R+4\\
& \hbox{or}\quad E=3R\\
& \hbox{or}\quad E=3R+3\\
\end{eqnarray}

Option 1) cannot be realized because such a state can only exist in a supermultiplet where the superconformal primary is a scalar in the fundamental representation of $SU(2)_R$ with energy $E=3/2$. This is a free field. One of its $R$-components is chiral, and so would give a positive contribution to the coefficient of $x$ in the index. The index does not contain such terms, hence there are no free chiral fields in the theory.

Option 2) cannot be realized because for the same reason: it cannot be a superconformal primary because there are no superconformal primaries in any unitary SCFT in the window $3/2<E<3$, and there is no free multiplet in our particular theories.

Thus, only the scalars with $E=3$, $j_1=0$, $R=1$ contribute to the second power of $x$. The conclusion is that in all the theories we consider there a superconformal multiplet whose superconformal primaries are scalars with energy $E=1$ which are in the adjoint representation of the $SU(2)_R$ group and are in the adjoint representation of $E_{N_f}$. More precisely, they are in the representation of $SO(2N_f)\times U(1)_I$ which is obtained from the adjoint representation of $E_{N_f+1}$ under restriction to the subgroup $SO(2N_f)\times U(1)_I\subset E_{N_f+1}$.

What remains to prove is that this is the superconformal multiplet of flavour currents. Flavour currents are primary operators with conformal dimension $\Delta=4$, they are $SU(2)_R$ singlets and are vectors under $Spin(5)$. In the radial quantization states with this quantum numbers can be obtained from scalars discussed in the previous paragraph by double action of supercharges as follows

\begin{equation}
\ket{vectors}\subset Q_{[\alpha}^{(i} Q_{\beta]}^{j)}\ket{scalars}
\end{equation}

The $SO(5)$ spinor indices $\alpha$ and $\beta$ of the $Q$'s are antisymmetrized to get an $SO(5)$ vector, and the $SU(2)_R$ indices are symmetrized because antisymmetrization produces the generator of translations making the vector descendant operators, while we need primaries.

However, this does not prove existence of the conserved currents because it may happen that the norm of these vectors is zero. Next I show that this does not occur.
Indeed, the norm is computed using (anti)commutation relations of the superconformal algebra, so it only depends on the representation of the primary operators under the bosonic subgroup of the superconformal group, it does not depend on the representation under the flavor group. So the norm of any representation of $SO(2N_f)\times U(1)_I$ is the same as the norm of the adjoint representation.

Next we ask in what superconformal multiplets the manifest currents of $SO(2N_f)\times U(1)_I$ live. The manifest flavor currents live in a superconformal multiplet where they cannot be superconformal primaries because, being flavor currents, they are annihilated by all supercharges $Q$'s, so they cannot be at the same time annihilated by all superconformal charges $S$'s$=Q^\dagger$'s in the radial quantization -- because of the anticommutation relations between $Q'$s and $S$s. Then by unitarity bounds their superconformal primaries are scalars with conformal dimension $E=3$ and either in a singlet or a triplet representation of $SU(2)_R$. However, they cannot be singlets because to get the currents from the scalars we would need to use the operator $Q_{\alpha}^{[i} Q_{\beta}^{j]}$. Then we have an option of either symmetrizing or antisymmetrizing the spinor indices of $SO(5)$. If we symmetrize them, we will not get a vector of $SO(5)$. If we antisymmetrize them, we will get the operators of momenta, so the vectors will not be primary.

So, the manifest flavor currents are obtained form scalars in the adjoint of $SU(2)_R$ with conformal dimension $E=3$ as

\begin{equation}
\ket{vectors}\subset Q_{[\alpha}^{(i} Q_{\beta]}^{j)}\ket{scalars}
\end{equation}

In other words, the norm of the state $\ket{vector}$ is positive but this is then true for the other representations of $SO(2N_f)\times U(1)_I$ that we have.

The conclusion is that in the supersymmetric 5-dimensional gauge theories with gauge group $SU(2)$ and $N_f=0,1,...7$ flavors in the fundamental representation there are conserved currents in the representation of the manifest flavor symmetry $SO(2N_f)\times U(1)_I$ which is obtained from the adjoint representation of $E_{N_f+1}$ under decomposition $SO(2N_f)\times U(1)_I\subset E_{N_f+1}$. Thus the current algebra forms the currents algebra of $E_{N_f+1}$ symmetry.  The global symmetry is enhanced to $E_{N_f+1}$.

As was noted above, two-instanton contributions to the indices were not computed in \cite{KKL}. This is irrelevant to our analysis for $N_f=0,1,...,5$ but not for $N_f=6,7$. Indeed, the decomposition of the adjoint representations of $E_7$ and $E_8$ under $SO(12)\times U(1)\subset E_{7}$ and $SO(14)\times U(1)\subset E_8$ are

\begin{eqnarray}
& 133=1_0+66_0+32_1+32_{-1}+1_2+1_{-2}\\
& 248=1_0+91_0+64_1+\overline{64}_{-1}+14_2+14_{-2}
\end{eqnarray}

By the above argument, only presence of conserved currents in representations $32_1+32_{-1}$ and $64_1+\overline{64}_{-1}$ in addition to the currents corresponding to the manifest flavor symmetries is guaranteed. However, the conserved current must form a closed algebra of some group. These groups have ranks $7$ and $8$ correspondingly because if there were any additional hidden $U(1)$, it showed up as additional term $+x^2$ in the indices. Furthermore, these groups are simple, so the only candidates are $A_7,B_7,C_7,D_7,E_7$ and $A_8,B_8,C_8,D_8,E_8$, correspondingly. There is a lower bound on their dimensions: $131$ and $220$, correspondingly. This bound excludes all options except $E_7$ and $E_8$. Thus, there is an enhancement of the global symmetry to $E_7$ and $E_8$ for $N_f=6$ and $N_f=7$, correspondingly.

\section{Acknowledgements}

 This work is supported by the Perimeter Institute for Theoretical Physics. Research at the Perimeter Institute is supported by the Government of Canada through Industry Canada and by Province of Ontario through the Ministry of Economic Development and Innovation.

\end{document}